\documentclass[floats,twocolumn,prl,shortbibliography,10pt]{revtex4-2}
\usepackage{amsfonts,amsmath,graphicx,epsfig,amssymb}
\usepackage{epsfig}
\usepackage{mathrsfs}
\usepackage{latexsym}
\usepackage{amsxtra}
\usepackage{amsbsy}
\usepackage{amscd}
\usepackage{color}
\usepackage{bbm}
\usepackage{hyperref}
\usepackage{multirow}
\usepackage{graphics}
\usepackage{amsfonts}
\usepackage{subfigure}
\usepackage{color}
\allowdisplaybreaks

\newcommand{\sectitle}[1]{\vspace{.5cm}{\em #1.--}}

\newcommand{\ee}{e^+e^-}

\newcommand{\BBb}{\overline YY}
\newcommand{\YYb}{\overline YY}
\newcommand{\ppb}{\overline pp}
\newcommand{\LLb}{\overline \Lambda\Lambda}
\newcommand{\LcLcb}{\overline \Lambda_c\Lambda_c}
\newcommand{\SSb}{\overline \Sigma\Sigma}
\newcommand{\LSb}{\overline \Lambda\Sigma^0}
\newcommand{\XXb}{\overline \Xi\Xi}
\newcommand{\XXbz}{\bar\Xi^0\Xi^0}
\newcommand{\XXbm}{\bar\Xi^+\Xi^-}
\newcommand{\OOb}{\overline \Omega\Omega}
\newcommand{\SSbp}{\bar\Sigma^-\Sigma^+}
\newcommand{\SSbz}{\bar\Sigma^0\Sigma^0}
\newcommand{\SSbm}{\bar\Sigma^+\Sigma^-}

\begin{document}

\title{Electromagnetic form factors of hyperons in the timelike region: A short review}

\author{Ling-Yun Dai$^{1,2}$}
\email{dailingyun@hnu.edu.cn}
\author{Johann Haidenbauer$^{3}$}
\email{j.haidenbauer@fz-juelich.de}
\author{Ulf-G. Mei{\ss}ner$^{4,3,5,6}$}
\email{meissner@hiskp.uni-bonn.de}
\affiliation{$^{1}$School for theoretical Physics, School of Physics and Electronics, Hunan University, Changsha 410082, China}
\affiliation{$^{2}$Hunan Provincial Key Laboratory of High-Energy Scale Physics and Applications, Hunan University, Changsha 410082, China}
\affiliation{$^{3}$Institute for Advanced Simulation (IAS-4), Forschungszentrum J\"ulich GmbH, D-52425 J\"ulich, Germany}
\affiliation{$^{4}$Helmholtz Institut f\"ur Strahlen- und Kernphysik and Bethe Center for Theoretical Physics, Universit\"at Bonn, D-53115 Bonn, Germany}
\affiliation{$^5$Peng Huanwu Collaborative Center for Research and Education, Beihang University, Beijing 100191, China}
\affiliation{$^{6}$Tbilisi State University,Tbilisi 0186, Georgia}

\begin{abstract}
We review recent experimental and theoretical results for the
electromagnetic form factors of hyperons (Y) in the timelike region, 
accessible in the reactions $e^+e^-\to \bar YY$. 
Specifically, we focus on the final states  
$\bar \Lambda\Lambda$, $\bar\Lambda\Sigma^0$/$\bar \Sigma^0\Lambda$, 
$\bar \Sigma\Sigma$, $\bar \Xi\Xi$, and $\bar\Omega\Omega$. 
The $\bar \Lambda_c\Lambda_c$ system is also discussed. 

\end{abstract}

\maketitle

\sectitle{Introduction}\label{Sec:I}
Electromagnetic form factors (EMFFs) constitute an important tool
for elucidating the internal structure of baryons. Indeed, the 
EMFFs of the proton (and with some restrictions of the neutron) in 
the spacelike region have been mapped out quite successfully 
by performing electron-proton ($ep$) scattering experiments and 
those measurements have provided clear evidence for the 
non-elementary nature of the nucleons. For recent reviews
see \cite{Denig:2012by,Pacetti:2014jai,Xia:2021agf,Lin:2021umz,Gao:2021sml}. 

For the hyperons ($Y$), that is the $\Lambda$, $\Sigma$, or $\Xi$, 
the situation is very 
different. Due to their short lifetime, preparing a hyperon target
is virtually impossible, and thus, electron scattering experiments
that allow access to the EMFFs in the spacelike region are not available. 
However, an experimental determination of the form factors is
possible in the timelike region, namely via the reaction $\ee\to\YYb$.
Some pioneering measurements have already been performed around 
two decades ago or even before \cite{DM2:1990tut,BaBar:2007fsu}. 
However, only recently has there been an enormous increase in the
wealth of data available for hyperons.
Not only an impressive variety of hyperons have 
been measured, from the $\Lambda$ up to the $\Omega$, 
but also a wide range of energies has been covered, reaching (in 
most cases) from close to the threshold at twice the hyperon mass
up to several GeV. 

In this review, we focus the development over the past few years, 
on the experimental as well as on the theoretical side.
The EMFFs in the timelike region are connected with those in the 
spacelike region by dispersion relations, which follow from
unitary and analyticity \cite{Denig:2012by,Lin:2021umz}. 
However, at the same time, they reflect different physical aspects 
of the baryons. As already mentioned above, in the spacelike
region, they provide information on the electromagnetic
structure of the baryons. In the timelike region, other
features can be studied. For example, close to the reaction
threshold the interaction of the produced $\BBb$ pair
should have a noticeable influence on the energy
dependence of the cross section and, in turn, 
on the EMFFs. Furthermore, the coupling of the photon to
vector mesons, exploited in the vector-meson dominance
(VMD) model, could be prominently seen in the cross section,
provided that there is also a non-zero coupling to the 
$\BBb$ state. 
At very high energies, one expects the onset 
of perturbative QCD (pQCD), as reflected in the energy
dependence that follows from the quark-counting rules. 
Besides, Ref.~\cite{Yang:2022qoy} suggests that the timelike EMFFs of the nucleons reflect the distributions of polarized electric charges induced by hard photons. This conjecture can be extended to the hyperons if it is confirmed. 

The review is structured in the following way:
In the next section, we provide the essentials of the formalism.
After that, we summarize the experimental situation and
 give a brief glimpse into the theoretical approaches.
Subsequently, a more detailed discussion of the cross sections 
and EMFFs for the $\Lambda$, $\Sigma^+$, $\Sigma^0$, $\Sigma^-$,
$\Omega$, and $\Lambda_c$, and of the $\Lambda\Sigma^0$ 
transition form factor is given, where the emphasis is on 
energies not too far from the thresholds. 
The review closes with a brief summary.

\sectitle{Formalism and observables}
Under the assumption that the reaction $\ee\to\YYb$ proceeds via 
one-photon exchange, the differential cross section and the
EMFFs $G^Y_E$ and $G^Y_M$ are related by
\begin{eqnarray}
\frac{d\sigma}{d\Omega} &=& \frac{\alpha^2\beta}{4 s}~C(s) 
\Big[\left| G^Y_M(s) \right|^2 (1+{\rm cos}^2\theta) \nonumber \\
&&+
\frac{4M_{Y}^2}{s} \left| G^Y_E(s) \right|^2 {\rm sin}^2\theta\Big] ~{\rm ,}
\label{eq:diff}
\end{eqnarray}
where $Y$ generically denotes the baryons 
$\Lambda$, $\Sigma$, $\Xi$, $\Omega$, and $\Lambda_c$,
and $\bar Y$ stands for the corresponding anti-particles.
Here, $\alpha = 1/137.036$ is the electromagnetic fine-structure constant and
$\beta=k_{Y}/k_e$ a phase-space factor, where $k_{Y}$ and $k_e$ are the
center-of-mass three-momenta in the $\YYb$ and $\ee$ systems, respectively,
related to the total energy via $\sqrt{s} = 2\sqrt{M_{Y}^2+k_p^2} = 2\sqrt{m_e^2+k_e^2}$.
Further, $m_e \, (M_{Y})$ is the electron (hyperon) mass.
For charged baryons, the Coulomb effect is present which is taken
into account via 
the $S$-wave Sommerfeld-Gamow factor $C(s)$, given by 
$C = y/(1-e^{-y})$ with $y = \pi \alpha M_{Y} /k_{Y}$.
In general, we omit the superscript $Y$ from the $G$'s in the following because it is anyway clear from the context in which hyperon is discussed.
As noted, the cross section as written in Eq.~({\ref{eq:diff}) results from the 
one-photon exchange approximation and by setting the electron mass $m_e$ to zero
(then $\beta = 2k_{Y}/\sqrt{s}$). In this case, the total
angular momentum is fixed to $J=1$ and the $\ee$ and $\YYb$ systems
can be only in the partial waves $^3S_1$ and $^3D_1$. 
Note further that the EMFFs in the timelike region are complex quantities.

The integrated reaction cross section is readily found to be
\begin{equation}
\sigma_{\ee \to \YYb}=\frac{4 \pi \alpha^2 \beta}{3s}~C(s)~
\left [ \left| G_M(s) \right|^2 + \frac{2M_{Y}^2}{s} \left| G_E(s) \right|^2 \right ].
\label{eq:tot}
\end{equation}

As is clear from Eqs.~(\ref{eq:diff}) and (\ref{eq:tot}) a separation of $G_E$ and $G_M$ is only possible from the differential cross section. Since just a few measurements of that observable have
been performed so far, another quantity is often 
considered, namely the effective baryon form factor 
$G_{\rm eff}$ which is defined by
\begin{equation}
|G_{\rm eff} (s)|=\sqrt{\sigma_{\ee\rightarrow \YYb} (s)\over {4\pi\alpha^2
\beta \over 3s} ~C(s)
\left [1 +\frac{2M^2_{Y}}{s}\right ]} \ .
\label{eq:Geff}
\end{equation}
This quantity provides a measure for the deviation of the
experimental cross section from the point-like one.

Finally, the relative phase of $G_E$ and $G_M$, defined by the
relation $G_E/G_M = \exp{(i\,\Delta\phi)}\, |G_E/G_M|$,
can be only determined from spin-dependent observables,
for example, it can be extracted from 
the analyzing power $A_y$, which is given
by \cite{Buttimore:2006mq}
\begin{equation}
A_y = \frac{\frac{2M_Y}{\sqrt{s}} \sin 2\theta \ {\rm Im}\, G^*_E(s) G_M(s)}
{
\left| G_M(s) \right|^2 (1+{\rm cos}^2\theta) +
\frac{4M_Y^2}{s} \left| G_E(s) \right|^2 {\rm sin}^2\theta}~{\rm .}
\label{eq:Ay}
\end{equation}

\sectitle{Overview of experiments} 
Cross section data are available for 
$\ee\to\LLb$
\cite{DM2:1990tut,BaBar:2007fsu,Dobbs:2014ifa,BESIII:2017hyw,BESIII:2019nep,BESIII:2023ioy},
$\ee\to\LSb$+c.c.
\cite{DM2:1990tut,BaBar:2007fsu,BESIII:2023pfv}, 
$\ee\to\SSb$ 
\cite{BaBar:2007fsu,Dobbs:2014ifa,
BESIII:2020uqk,BESIII:2021rkn,BESIII:2023ynq,BESIII:2023ldb,BESIII:2024umc,
Belle:2022dvb}, 
$\ee\to\XXb$
\cite{Dobbs:2014ifa,BESIII:2020ktn,BESIII:2021aer,BESIII:2023rse,
BESIII:2024ues}, 
$\ee\to\OOb$ 
\cite{Dobbs:2014ifa,BESIII:2022kzc}, and 
$\ee\to\LcLcb$
\cite{Belle:2008xmh,BESIII:2017kqg,BESIII:2023rwv},  
For an overview of data from the BESIII 
Collaboration see also \cite{Schonning:2023mge}.
%
One essential advantage of hyperons is that the 
self-analyzing character of their weak decay allows one 
to determine also the polarization 
as well as spin-correlation parameters. 
With those observables one can determine the relative phase
between $G_E$ and $G_M$~\cite{Buttimore:2006mq}, as already
mentioned. 
However, so far, detailed information like the aforementioned
polarizations or differential cross sections
is only available for a few $\bar BB$ channels,
namely $\LLb$ \cite{BaBar:2007fsu,BESIII:2019nep}, 
$\SSbp$ \cite{BESIII:2020uqk,BESIII:2023ynq}, 
and $\LcLcb$
\cite{BESIII:2017kqg,BESIII:2023rwv}. 

\sectitle{Theoretical approaches}
At low energies, the energy dependence of the $\ee\to\BBb$ 
cross section should be strongly influenced by the final-state
interactions (FSI) in the $\BBb$ system. This has been 
observed for protons \cite{Haidenbauer:2014kja} (and references therein) 
and it is likewise expected for hyperons. Therefore, such data
constitute a testing ground and source of information on 
the interaction between the produced $\BBb$ pair. 
The information is complementary to the one from $\ppb$ induced 
production of $\YYb$ systems studied with the LEAR facility at 
CERN~\cite{Klempt:2002ap}. FSI effects due to $\bar NN$ and 
$\YYb$ interactions
have been the focus of our studies~\cite{Kang:2015yka, Haidenbauer:2016won,Haidenbauer:2020wyp,Dai:2017fwx,Guo:2024pti}
and those of others too, see, e.g. 
\cite{Loiseau:2005cv,Entem:2007bb,Dalkarov:2009yf,Chen:2010an,Salnikov:2023azb,
Milstein:2022bfg,Salnikov:2023qnn,Jia:2024ybo}. 

The formalism for the inclusion of the FSI in the
$\BBb$ systems applied by us is described in detail in
Refs.~\cite{Haidenbauer:2014kja,Haidenbauer:2016won}.
Here, we restrict ourselves to an illustration; see Fig.~\ref{fig:FSI}. 
$G^0_M$ and $G^0_E$ represent the bare vertices or bare 
form factors, respectively, which we assume to be 
constants. Since $G^0_M = G^0_E$ at the reaction 
threshold due to kinematical constraints 
\cite{Haidenbauer:2014kja}, there is, in general, only 
a single (complex) parameter that enters the calculation.
In practice, it is a normalization factor to be fixed from 
the measured cross sections. 
The physical (dressed) EMFFs $G_E$ and $G_M$ acquire their
energy (momentum) dependence in their evaluation in distorted
wave Born approximation, based on a $\BBb \to\BBb$ 
amplitude (half-off-shell $T$-matrix) which we 
take from our analyses of the reactions $\ppb\to\LLb$,
$\LSb$+c.c., $\SSb$, $\XXb$ that were measured in
the PS170 experiment at the LEAR facility at 
CERN~\cite{Klempt:2002ap}. The intermediate
states $\bar Y'Y'$ are restricted to
$\LLb$ ($\LSb$/$\bar\Sigma^0\Lambda$) in the reactions 
$\ee\to\LLb$ ($\ee\to\LSb$+c.c.), and consist of 
$\SSbp$, $\SSbz$, $\SSbm$ for the three
$\ee\to\SSb$ channels, and of $\XXbz$ and
$\XXbm$ for the two $\ee\to\XXb$ channels. 

\begin{figure}[htbp] 
\begin{center}
\includegraphics[width=0.95\linewidth]{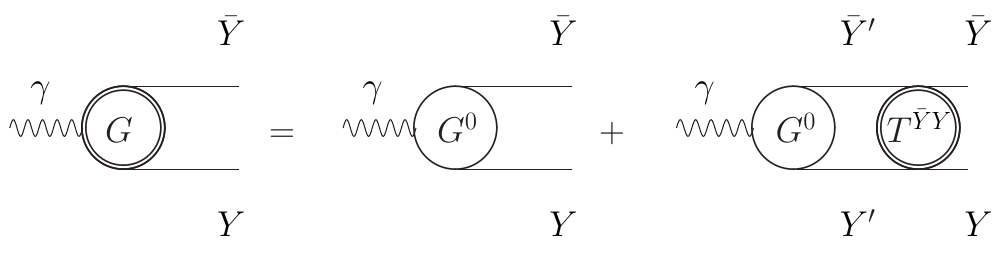}
\end{center}
\vskip -0.6cm
\caption{Illustration of the treatment of the FSI in 
the reaction $\ee\to\YYb$. Here, $G^0$ denotes  the 
bare EMFFs and $T^{\YYb}$ is the $\bar Y'Y'\to \YYb$
reaction amplitude. 
}
\label{fig:FSI}
\end{figure} 

In the VMD model, the virtual photon couples to the
baryons through vector mesons~\cite{Sakurai:1960ju,Meissner:1987ge,Pacetti:2014jai}. 
The standard vector mesons included in such studies are the
$\rho$, the $\omega$, and also the $\phi$. However, their poles lie 
in the unphysical region and, in particular, well below the $\BBb$
thresholds, so that there is only a moderate effect on the energy
dependence of the $\ee\to\BBb$ cross section. This is
different for the $\phi(2170)$ whose pole is located fairly close 
to the $\LLb$ threshold. See Ref.~\cite{Lorenz:2015pba} for a discussion
in the $\bar{p}p$ system.
A prominent effect is also expected from the
$\psi(3770)$ resonance. In actual parameterizations of the $\ee\to\BBb$
amplitudes within the VMD model, additional vector mesons 
are usually introduced. 
The imaginary part of the EMFFs is generated by including the finite 
width of the vector mesons. Finally, in general, a so-called intrinsic
form factor of dipole form, $g(s) = 1 / (1-\gamma s)^2$, is 
multiplied where the parameter $\gamma$ is adjusted to the data.  

At sufficiently high energies one expects to see the onset of
pQCD \cite{Brodsky:1974vy,Lepage:1979za},
characterized by an energy dependence that follows from the 
so-called quark counting rules, 
see also the review \cite{Pacetti:2014jai}.
A QCD-inspired model for the EMFFs of hyperons with 
focus on the perturbative regime of QCD 
has been proposed by Ramalho et al. \cite{Ramalho:2019koj,Ramalho:2020laj,Ramalho:2024wxp}. It is
an extension of their covariant spectator quark model in the 
spacelike regime to the timelike region \cite{Ramalho:2019koj}. 
In this context, we should also mention
the simple pQCD-inspired parameterization 
by \cite{Yang:2017hao} and the one utilized in various
papers by BESIII, taken from \cite{Pacetti:2014jai},
which reproduce the energy dependence of the $\YYb$ EMFFs 
in the timelike region reasonably well, except for the
threshold region. 

Discussions of the hyperon form factors in the 
timelike region from other perspectives can 
be found in~\cite{Bianconi:2022yjq,Dai:2023vsw}.

For completeness, let us mention that spacelike EMFFs of hyperons
are considered in \cite{Lin:2022dyu,Liu:2023reo}. See also
\cite{Kubis:2000aa,Yang:2020rpi}.

\sectitle{$\ee\to\LLb$ and the EMFFs of the $\Lambda$}
%
Here, new cross section
data have become available last year from BESIII~\cite{BESIII:2023ioy}.
The data in the near-threshold
region are presented in Fig.~\ref{fig:LL}, together
with some theory results from the literature. 
The predictions from Refs.~\cite{Haidenbauer:2016won,Haidenbauer:2020wyp}
are based on FSI effects of $\LLb$ potentials established in 
studies of the reaction $\ppb\to\LLb$~\cite{Haidenbauer:1991kt,Haidenbauer:1993ws}. 
In fact, the reaction $\ppb\to\LLb$ has
been extensively investigated in the PS185 experiment, and data 
are available for total and differential cross sections
as well as for spin-dependent observables, down to energies 
very close to the reaction threshold, see e.g., the
review in Ref.~\cite{Klempt:2002ap}. 
Thus, those $\LLb$ potentials are fairly well constrained by data.

\begin{figure}[t] 
\begin{center}
\includegraphics[width=0.95\linewidth]{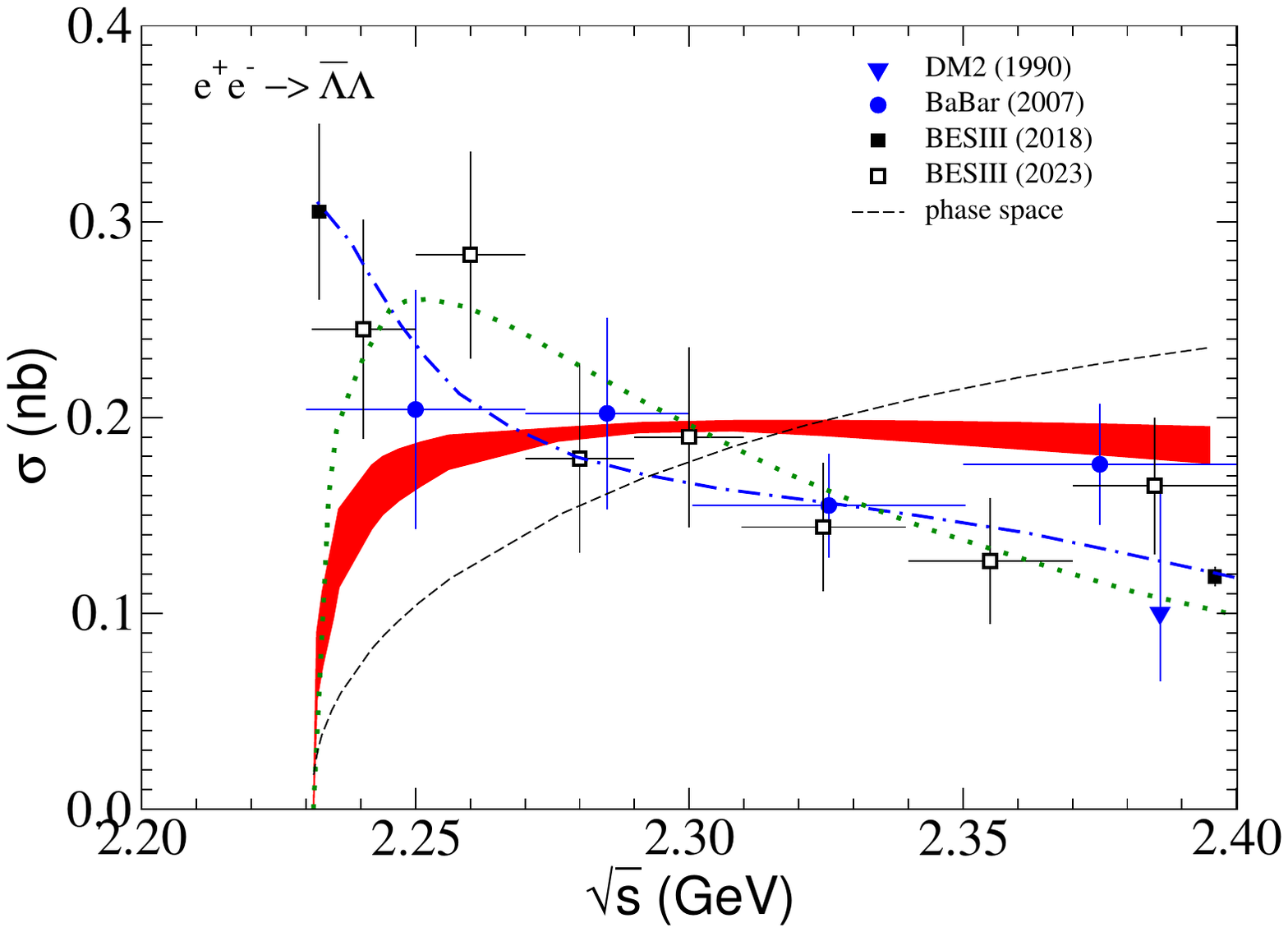} 
\includegraphics[width=0.95\linewidth]{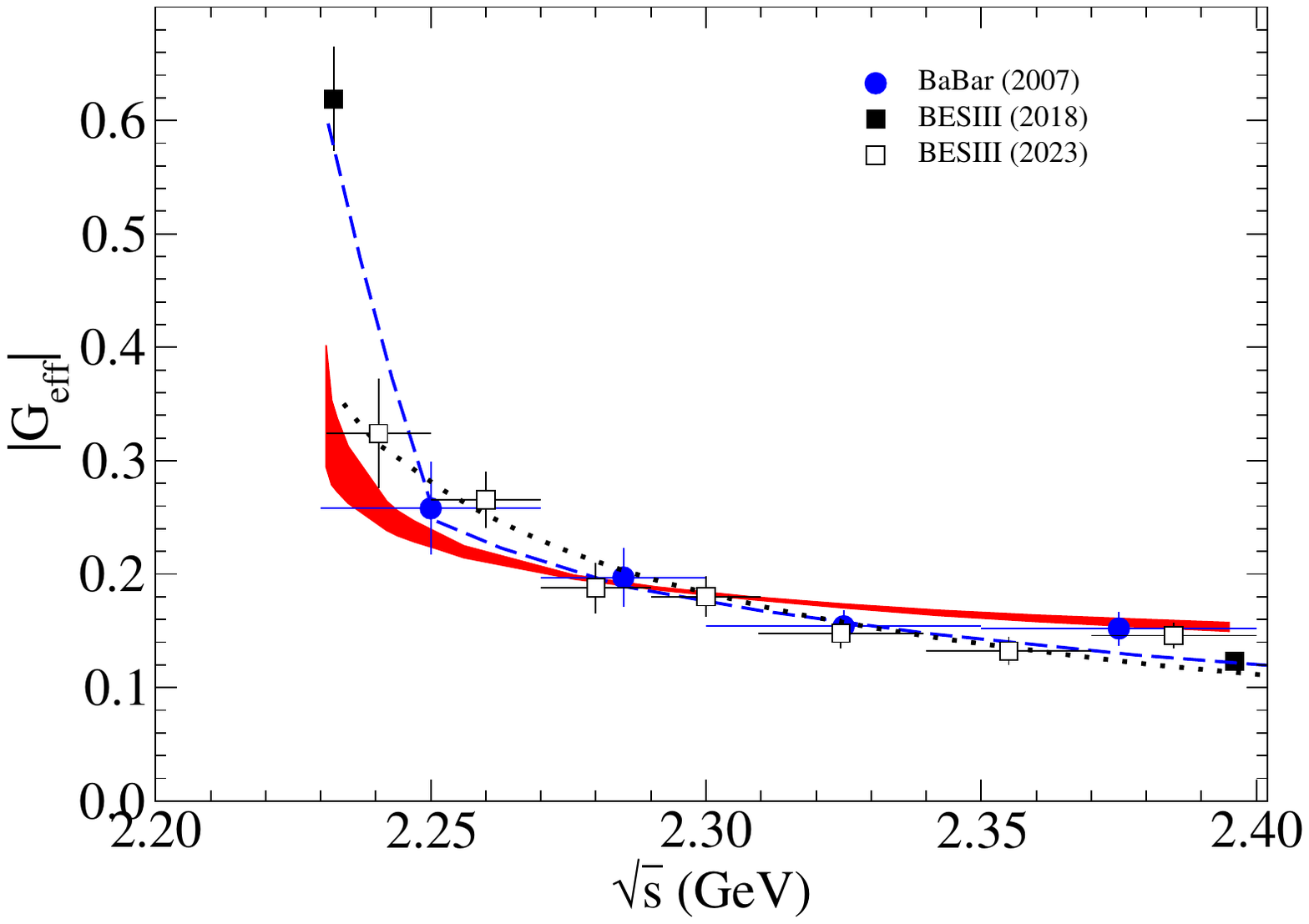}
\end{center}
\vskip -0.8cm 
\caption{Cross section for $\ee\to\LLb$ and pertinent $G_{\rm eff}$. 
Results from Refs.~\cite{Haidenbauer:2016won,Haidenbauer:2020wyp} 
based on the $\LLb$ FSI fixed in \cite{Haidenbauer:1993ws} are shown 
by the red band, 
while the dotted line is the FSI fit from~\cite{Salnikov:2023azb}. 
The dashed-dotted and dashed lines are VMD results from~\cite{Cao:2018kos}
and \cite{Li:2021lvs}, respectively. 
}
\label{fig:LL}
\end{figure}

\begin{figure}[t] 
\begin{center}
\includegraphics[width=0.92\linewidth]{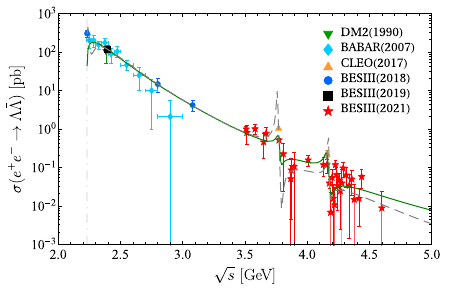}
\end{center}
\vskip -0.8cm 
\caption{Cross section for $\ee\to\LLb$. 
 Results of a dispersion-theoretical analysis by Lin et al. taken from~\cite{Lin:2022baj}.
}
\label{fig:LLu}
\end{figure}

The results are shown as bands that represent the model
uncertainties. Obviously, the $\LLb$ FSI predicts a basically 
flat behavior of the cross section, quite in line with the 
energy dependence of the BaBar data \cite{BaBar:2007fsu}.
On the other hand, the BESIII data point from 
2018 at $2232.4$~MeV~\cite{BESIII:2017hyw}
(filled square), i.e. just about $1$~MeV above the $\LLb$ threshold
($2M_\Lambda = 2231.37$~MeV), deviates from the overall trend and is 
not reproduced by the calculation.  
The already mentioned new BESIII data from 2023~\cite{BESIII:2023ioy} 
(open circles) are
consistent with the other measurements in the threshold region 
within the uncertainties.
In addition, they seem to
support a tendency towards a more regular threshold behavior. 
Indeed, it has been shown that the data 
can be described by a conventional FSI \cite{Salnikov:2023azb},
cf. the dotted line, 
though it remains unclear whether in that case the required $\LLb$ 
interaction is compatible with constraints from the $\ppb\to\LLb$ data. 

Data for $|G_E/G_M|$ and $\Delta \phi$ are available at a
few energies \cite{BaBar:2007fsu,BESIII:2019nep}
and are discussed in~\cite{Haidenbauer:2020wyp}.

Results for $\LLb$ within the VMD approach have been presented 
in Refs.~\cite{Yang:2019mzq,Xiao:2019qhl,Li:2021lvs,Bai:2023dhc}.
Besides the standard $\omega$ and $\phi$ contributions the 
$\omega(1420)$, $\omega(1650)$, $\phi(1680)$, $\phi(2170)$ have 
been included in~\cite{Yang:2019mzq}. 
In Ref.~\cite{Bai:2023dhc}
various broad $\phi(nS)$ and $\phi(nD)$ states with masses between
$2423$ and $2924$~MeV have been taken into account. A similar 
plethora of states has also been considered in \cite{Xiao:2019qhl}.
The work of Li et al.~\cite{Li:2021lvs} explores specifically 
the unusual behavior at the $\LLb$ threshold. 
It demonstrates that the description of the BESIII close-to-threshold data point 
requires to add a narrow and so far unknown resonance whose
mass coincides with the $\LLb$ threshold within its width
($M_x = 2230.9$~MeV, $\Gamma_x = 4.7$~MeV). 
The behavior at the threshold has also been studied
by Cao et al.~\cite{Cao:2018kos}.
In this case, a background motivated by pQCD is adapted, supplemented
with a suitably adjusted contribution of the sub-threshold resonance 
$\phi(2170)$ in combination with an additional resonance at 
$2340$~MeV. The results of the latter calculations are reproduced in 
Fig.~\ref{fig:LL} for illustration, cf. the dash-dotted line.

An exemplary result of what can be achieved within the VMD approach 
over a larger energy region is shown in Fig.~\ref{fig:LLu}. 
This calculation is taken from Ref.~\cite{Lin:2022baj} where a 
dispersion-theoretical analysis
of the $\Lambda$ EMFFs in the timelike region has been performed.  
Here, several narrow and also broad vector mesons (poles) in the mass region of $1965$~MeV to $4170$~MeV are included when fitting to the $\ee\to\LLb$ data, among others, the $\phi(2170)$ and the $\psi(3770)$. 
One can see that the cross section can be rather 
well described up to $\sqrt{s}\approx 5$~GeV.
Further studies of $\ee\to\LLb$ can be found in \cite{Baldini:2007qg,Dalkarov:2009yf,Yang:2019mzq}.

\begin{figure}[htbp] 
\begin{center}
\includegraphics[width=0.95\linewidth]{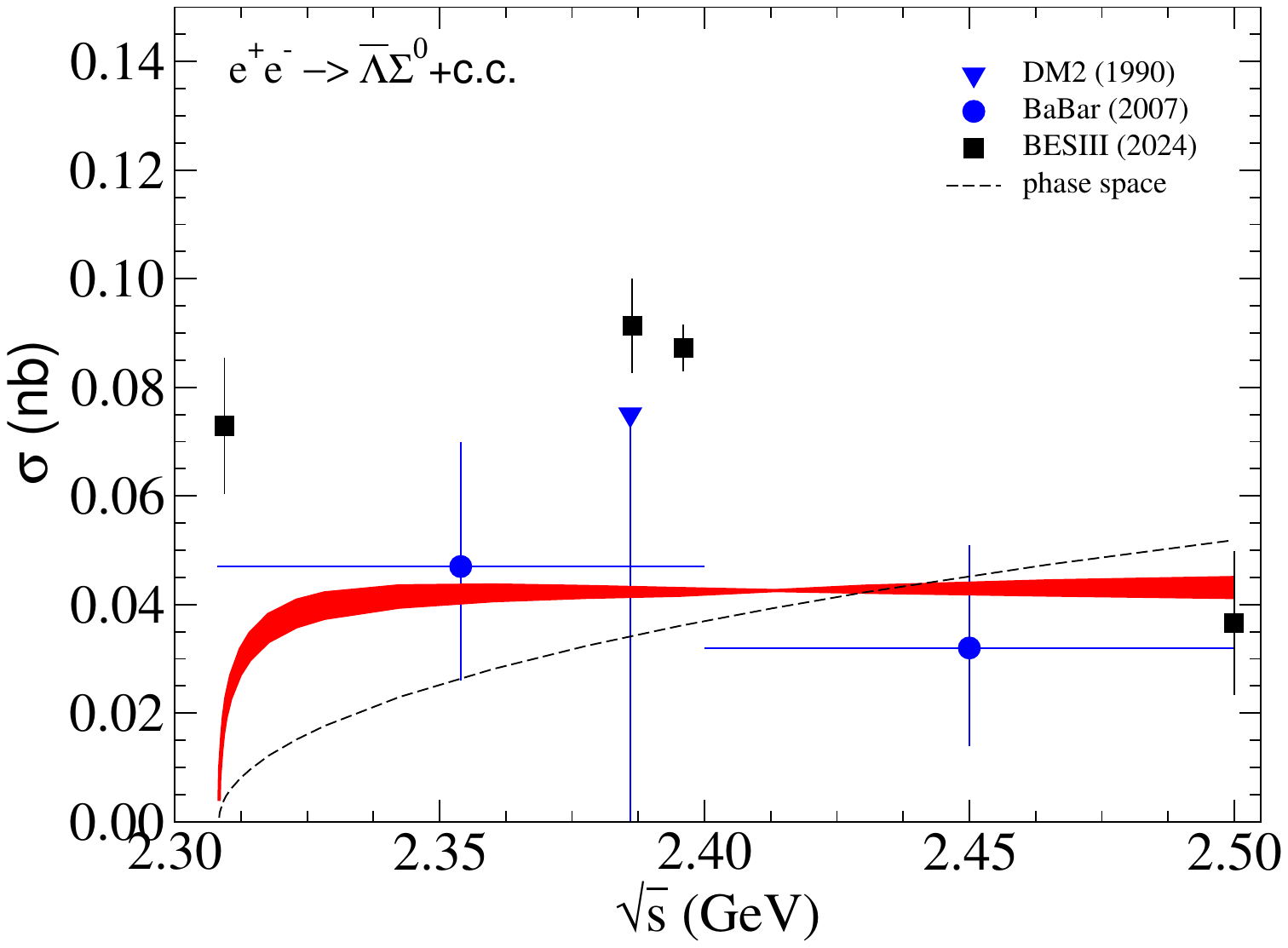}
\includegraphics[width=0.95\linewidth]{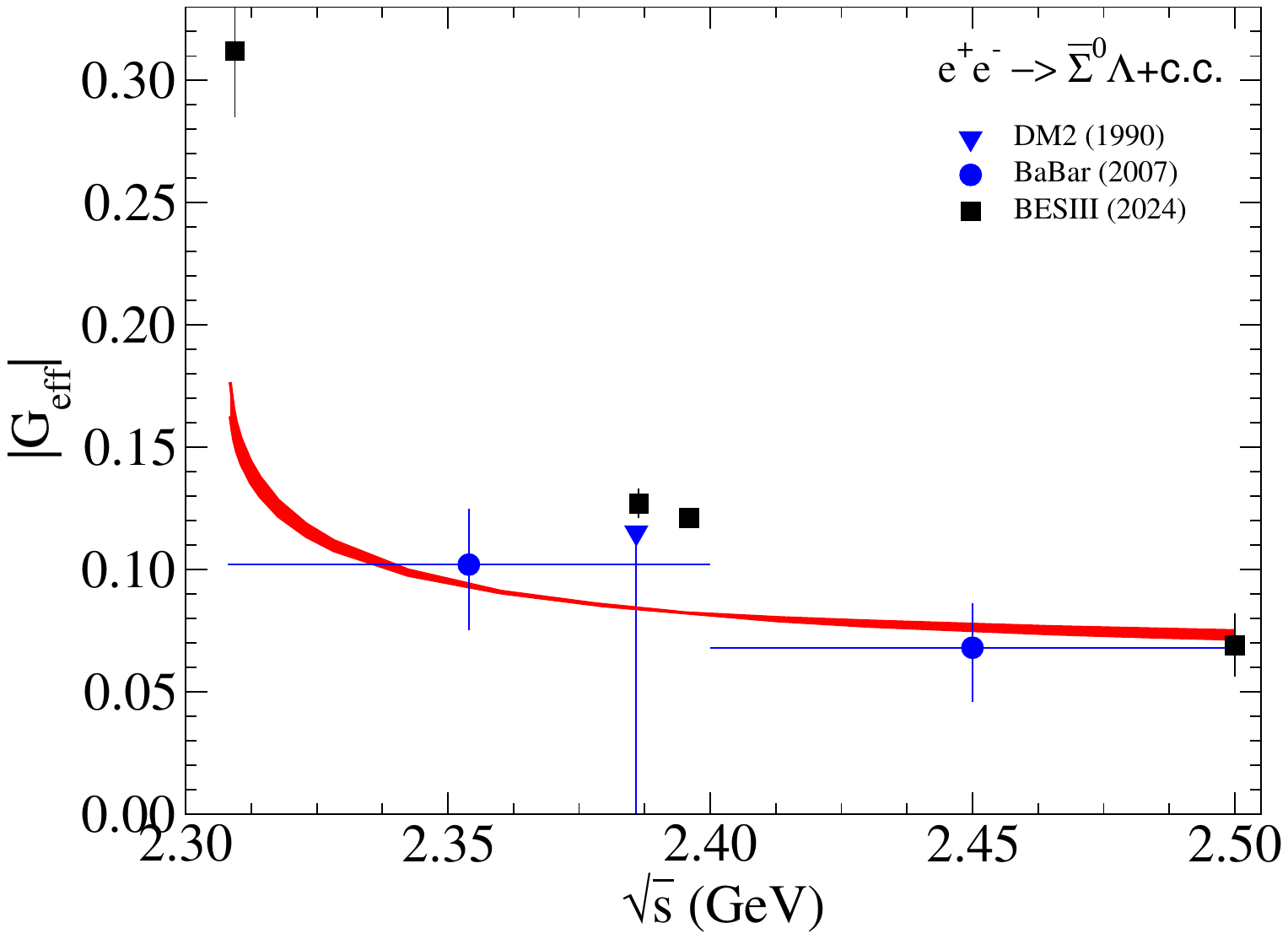}
\end{center}
\vskip -0.8cm 
\caption{Cross section for $\ee\to\LSb$+c.c. and pertinent 
$G_{\rm eff}$.
Data are from DM2~\cite{DM2:1990tut},
BaBar~\cite{BaBar:2007fsu}, and BESIII
\cite{BESIII:2023pfv}.
The theory result (band) is from~\cite{Haidenbauer:2020wyp}.
}
\label{fig:LS}
\end{figure} 

\sectitle{$\ee\to\LSb$+c.c. and the transition EMFFs for $\Lambda\Sigma^0$}
For the reaction $\ee \to \LSb$+c.c., so far, only cross sections 
are available. Pertinent results are presented in Fig.~\ref{fig:LS}.
Recently, BESIII has provided new data~\cite{BESIII:2023pfv}
(squares). Previously, only two measurements had been
available~\cite{DM2:1990tut,BaBar:2007fsu} where the one 
from DM2 \cite{DM2:1990tut} is actually an upper limit.

Before we start discussing the FSI effects, let us emphasize 
that, compared to $\ppb\to\LLb$, the experimental situation for the 
reaction $\ppb \to\LSb$+c.c. 
is much less satisfactory, and for $\ppb\to\SSb$ basically only 
a single measurement exists \cite{Klempt:2002ap}. 
In case of $\ppb\to \XXb$, only
upper bounds for the reaction cross section are available. 
Thus, the quality and quantity
of constraints on the $\YYb$ interactions that have been applied in~\cite{Haidenbauer:2020wyp} to obtain the results for the
$\ee\to\YYb$ channels in question are much more limited. 
This should be kept in mind. 
Also, note that
in Ref.~\cite{Haidenbauer:2020wyp} the normalization (i.e., the bare 
form factors $G^0_E$, $G^0_M$) have been fixed to the BaBar data, 
so that the cross sections are $0.04$~nb at $2.4$~GeV. 

Overall, the $\LSb$+c.c. cross section is noticeably  
smaller than that for $\LLb$.
However, like the latter, it remains practically constant over the energy region considered and
does not follow the phase-space behavior, at least according
to the BaBar measurement ~\cite{BaBar:2007fsu}. 
The predictions based on the $\YYb$ models established in
Ref.~\cite{Haidenbauer:1993ws} agree nicely with that 
behavior. The new data from BESIII~\cite{BESIII:2023pfv} suggest
a near-threshold cross section that is almost twice as large and
possibly more energy-dependent, too. 

\begin{figure}[ht] 
\begin{center}
\includegraphics[width=0.91\linewidth]{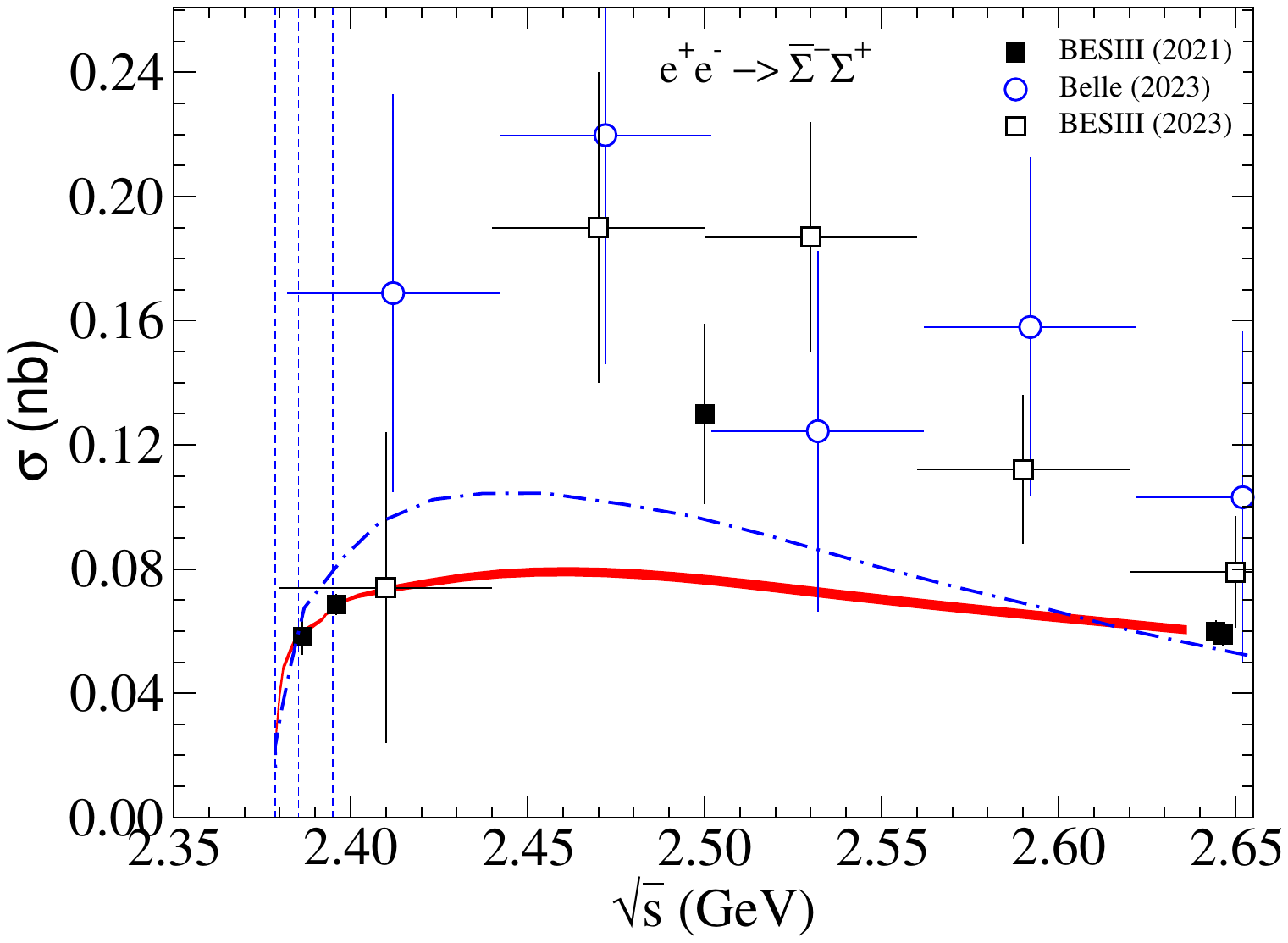}
\includegraphics[width=0.91\linewidth]{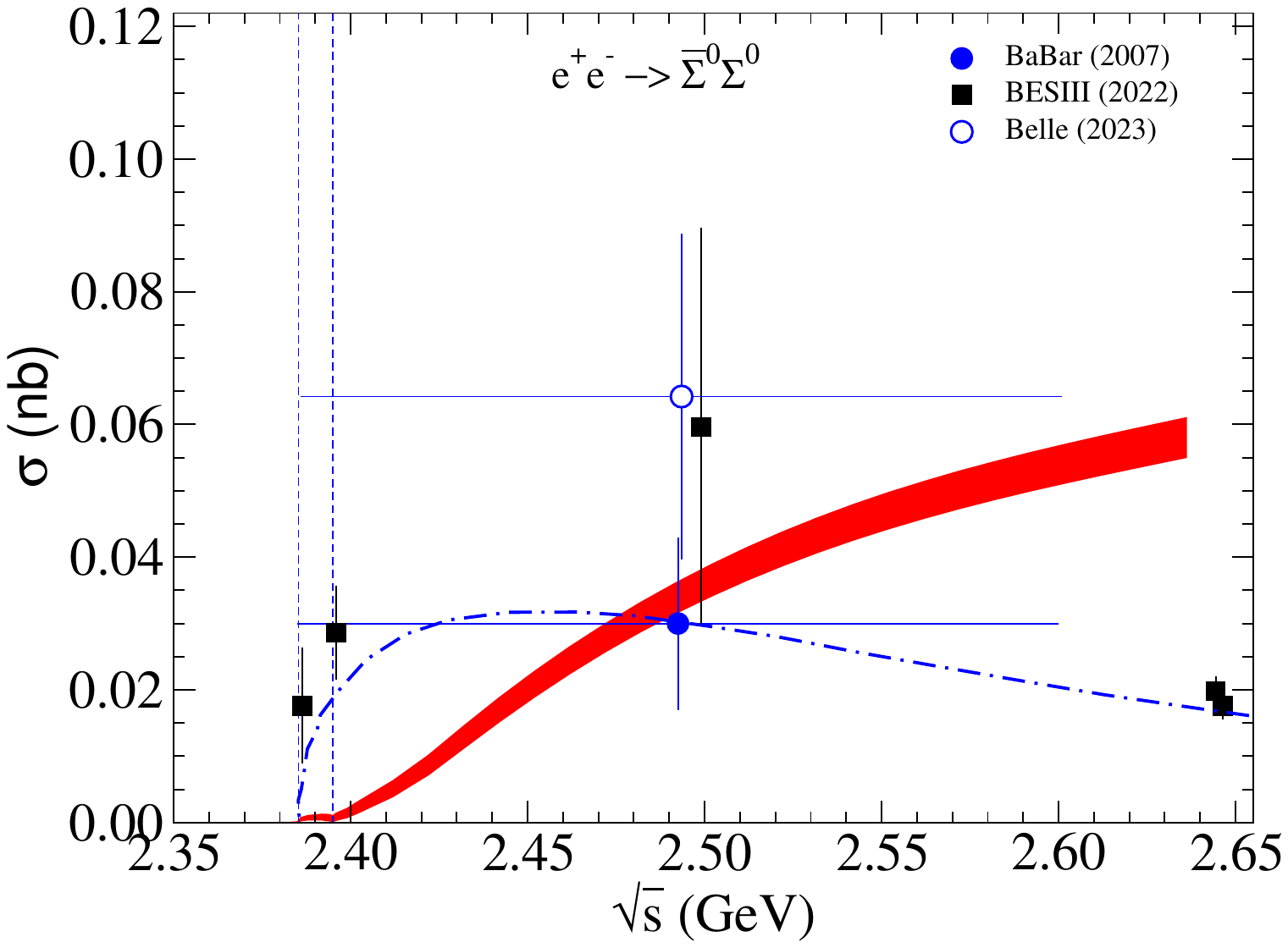}
\includegraphics[width=0.91\linewidth]{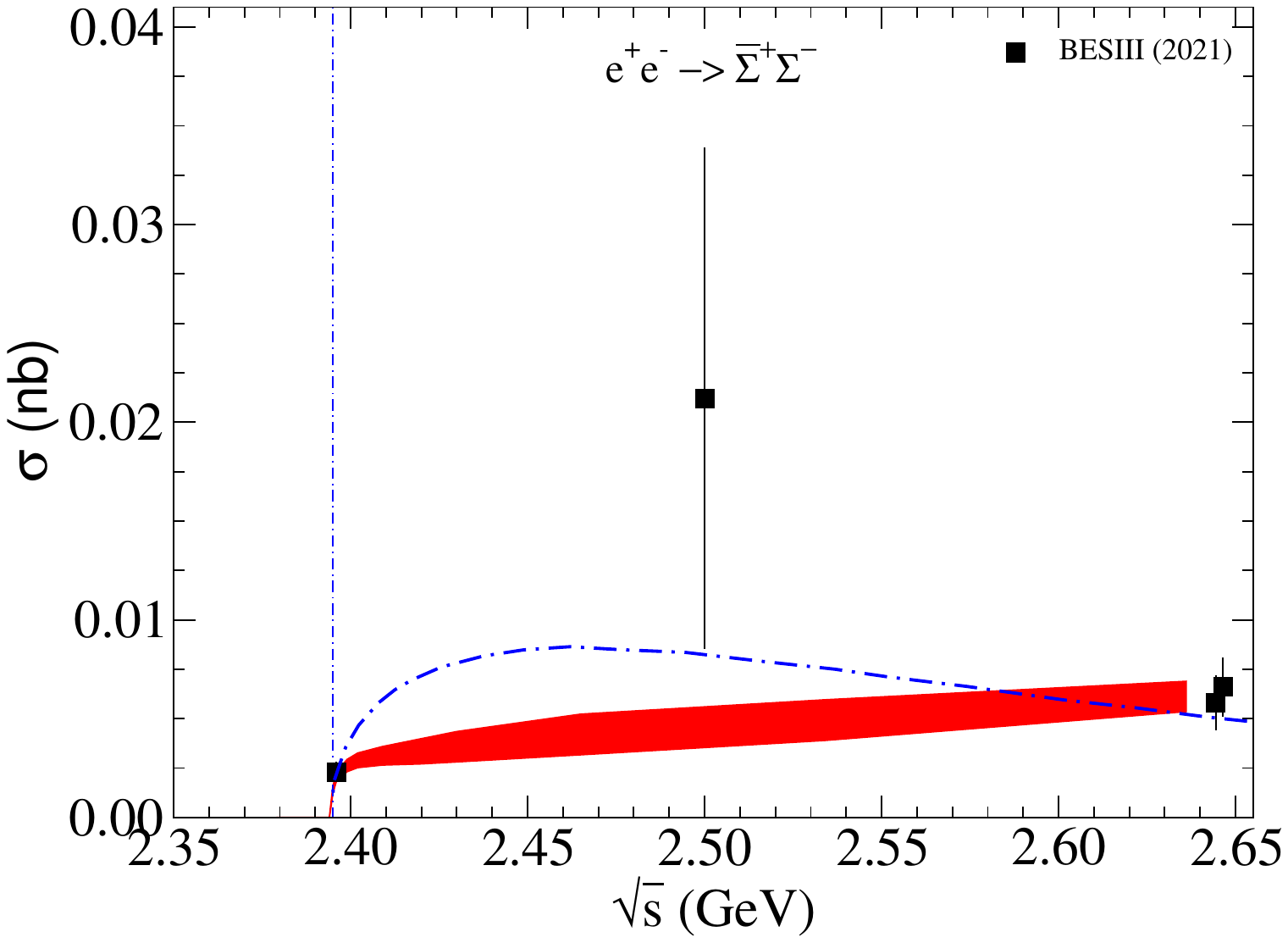}
\end{center}
\vskip -0.6cm 
\caption{Cross sections for $\ee\to\SSbp,\, \SSbz,\, \SSbm$
(top left, right, bottom). 
Data are from BaBar \cite{BaBar:2007fsu},
BESIII (2020) \cite{BESIII:2020uqk},  
(2021) \cite{BESIII:2021rkn}, and
(2023)~\cite{BESIII:2023ldb}, 
and Belle~\cite{Belle:2022dvb}.
Predictions from~\cite{Haidenbauer:2020wyp} based on
FSI effects are shown by bands. 
The dash-dotted line is the VMD result from \cite{Yan:2023yff}.
Vertical lines indicate the $\SSb$ thresholds. 
}
\vskip -0.4cm
\label{fig:SS}
\end{figure}

\sectitle{$\ee\to\SSb$ and the EMFFs of the $\Sigma$}
Here, new measurements have been reported by the
Belle Collaboration~\cite{Belle:2022dvb} 
($\SSbp$ and $\SSbz$ cross sections), 
and by BESIII ($\SSbp$ cross sections~\cite{BESIII:2023ldb} 
and values for $|G_E/G_M|$ and $\Delta \phi$~\cite{BESIII:2023ynq}).
An overview of the situation near the threshold is
provided in Figs.~\ref{fig:SS} and \ref{fig:SSFF}.

As already emphasized above, the rather limited information 
on the $\ppb\to\SSb$ reaction~\cite{Klempt:2002ap} did not allow 
to constrain the $\SSb$ interaction in~\cite{Haidenbauer:1993ws} 
reliably. 
Thus, the $\ee\to\SSb$ results of \cite{Haidenbauer:2020wyp},
shown as bands in Fig.~\ref{fig:SS}, 
have primarily an exploratory character. 
Nonetheless, it turned out that the energy dependence on the
channels $\ee \to \SSbp, \SSbm$, where data at low energies 
were available at the time when that study was performed,  
can be roughly described. Moreover, it was found that 
the $\SSb$ FSI introduces a strong interplay between the 
$\ee\to\SSbp, \SSbz, \SSbm$ results in the near-threshold region.

\begin{figure}[htbp] 
\begin{center}
\includegraphics[width=0.95\linewidth]{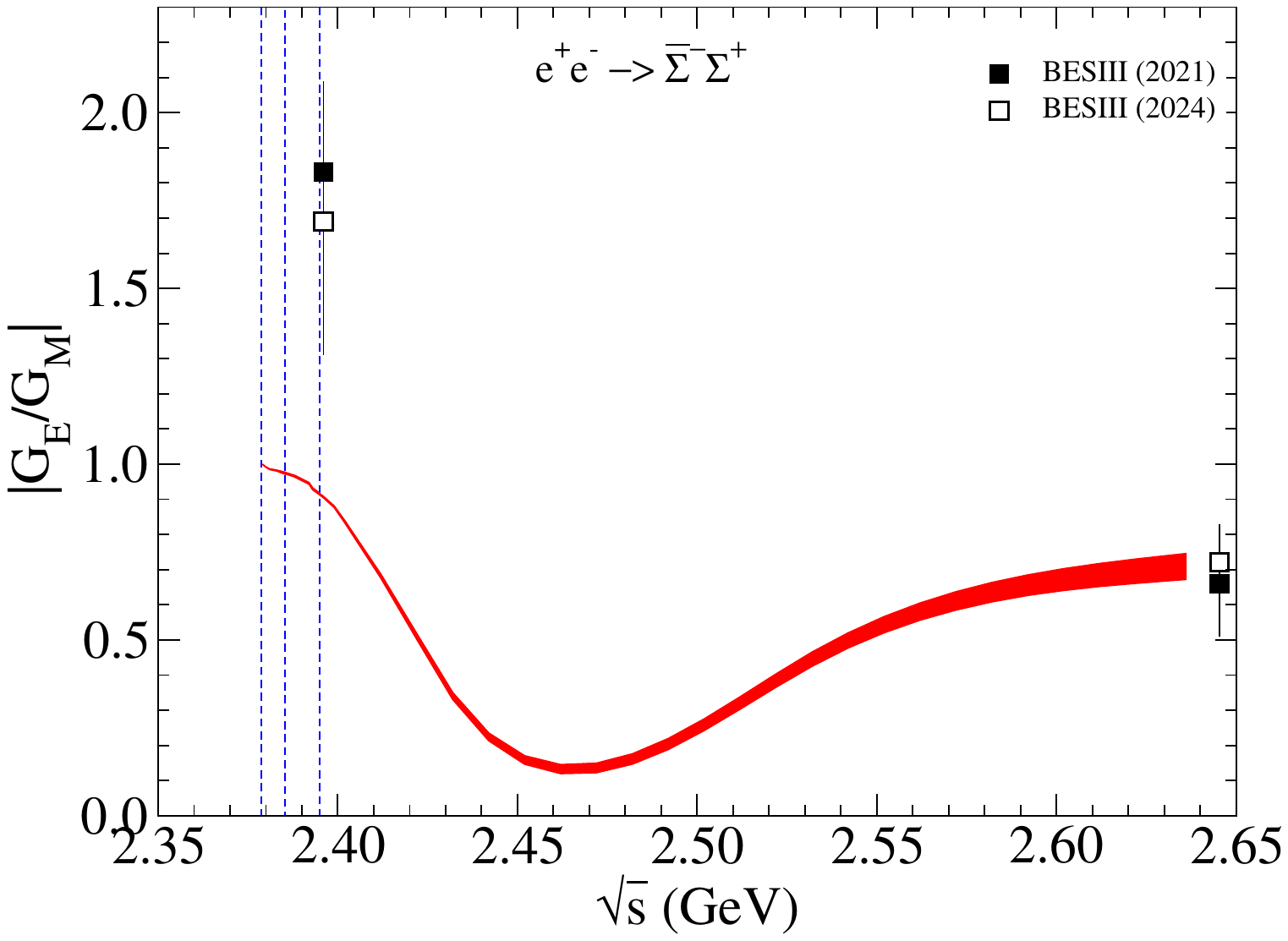}
\includegraphics[width=0.95\linewidth]{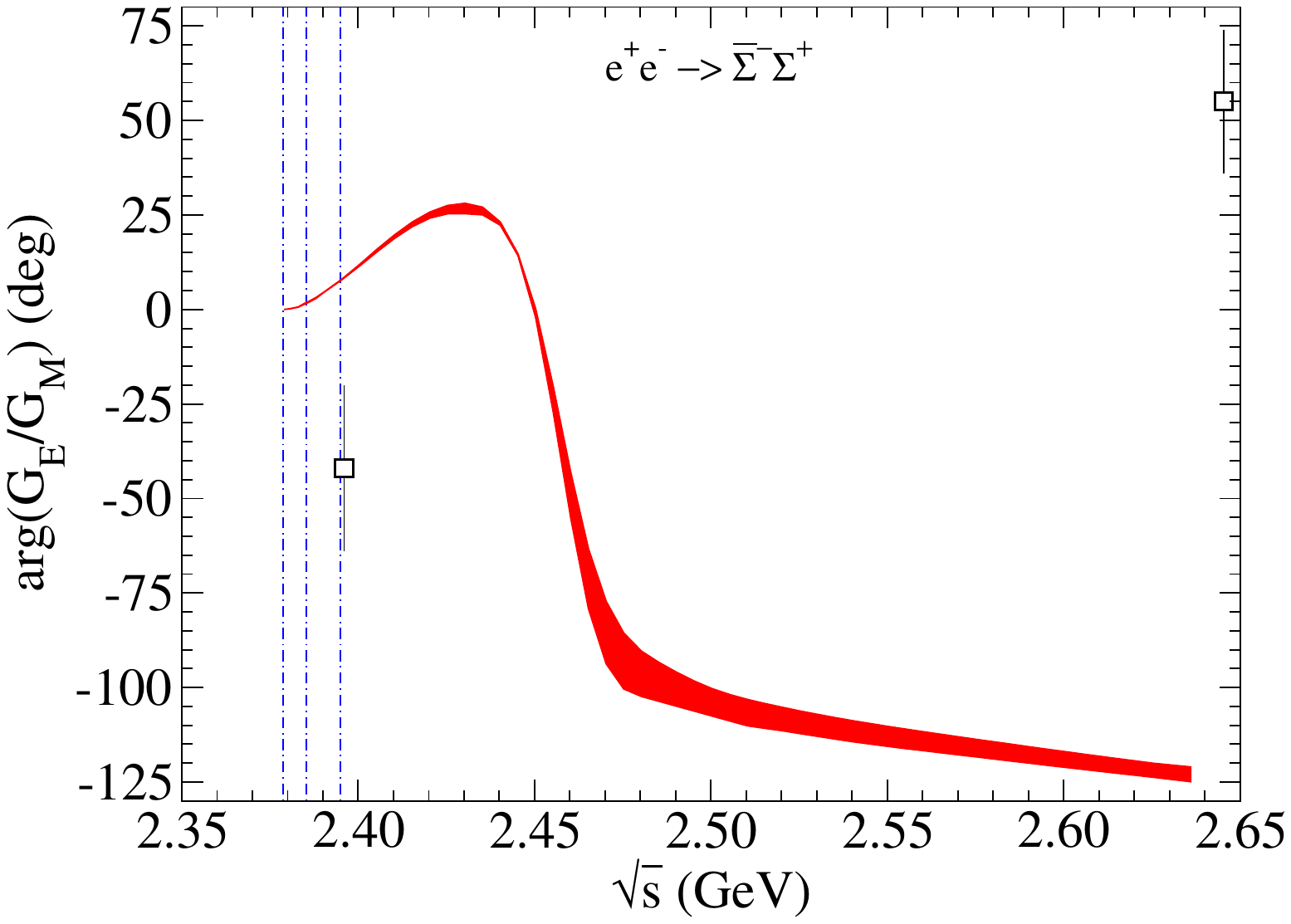}
\end{center}
\vskip -0.8cm 
\caption{Form factor ratio $|G_E/G_M|$ and
phase $\Delta \phi$ for $\ee\to\SSbp$. 
Data are from \cite{BESIII:2020uqk,BESIII:2023ynq}.
The theory results are from~\cite{Haidenbauer:2020wyp}.
Vertical lines indicate the $\SSb$ thresholds. 
}
\label{fig:SSFF}
\end{figure} 

The new data for $\SSbp$ \cite{Belle:2022dvb,BESIII:2023ldb} 
suggest a cross section significantly larger than 
what had been reported before while those for $\SSbz$
\cite{BESIII:2021rkn,Belle:2022dvb} indicate that
the energy dependence here could be similar to that 
in the other $\SSb$ channels. 
Indeed, to some extent, the experimental situation for $\SSbp$
has become confusing when one compares the earlier 
high-precision measurements by BESIII \cite{BESIII:2020uqk} 
(filled squares) very close to the threshold with the overall 
trend of the other data. 
Hopefully, future experiments will allow one to 
establish reliably the energy dependence of the $\SSbp$
cross section at low energies, but also the ones of the 
other $\SSb$ channels. The same applies to other observables
like the ratio $|G_E /G_M |$ or the relative phase between $G_E$ and
$G_M$. The latter, so far, measured only for 
$\SSbp$ ~\cite{BESIII:2023ynq}, seem to be very 
sensitive to the details of the $\SSb$ interaction as one can conclude 
from Fig.~\ref{fig:SSFF}. But concrete conclusions on the $\SSb$
dynamics are difficult to draw from just two data points.
In any case, since $G_E=G_M$ at the threshold, large
values for $|G_E/G_M|$ and $\Delta \phi$ so close to the threshold,
as indicated by the BESIII experiment are difficult to
achieve with a regular $\YYb$ interaction. 
Most likely, it would require again the introduction of 
a hypothetical resonance. 


Calculations for the EMFFs of the $\Sigma$ within the VMD approach 
can be found in two publications~\cite{Yan:2023yff,Li:2020lsb}, 
where in both works $\rho$, $\omega$, and $\phi$ are 
taken as the starting point. 
Results for all $\Sigma$ channels have been reported in 
Ref.~\cite{Yan:2023yff}. Those are included in Fig.~\ref{fig:SS}; see the dash-dotted lines. As one can see, with that approach 
the energy dependence in all three channels can be reproduced 
fairly well.
Results for $\Sigma^+$ and $\Sigma^-$ can be found in Ref.~\cite{Li:2020lsb}, and those show a comparable 
agreement with the empirical information. 

\begin{figure}[htbp] 
\begin{center}
\includegraphics[width=0.95\linewidth]{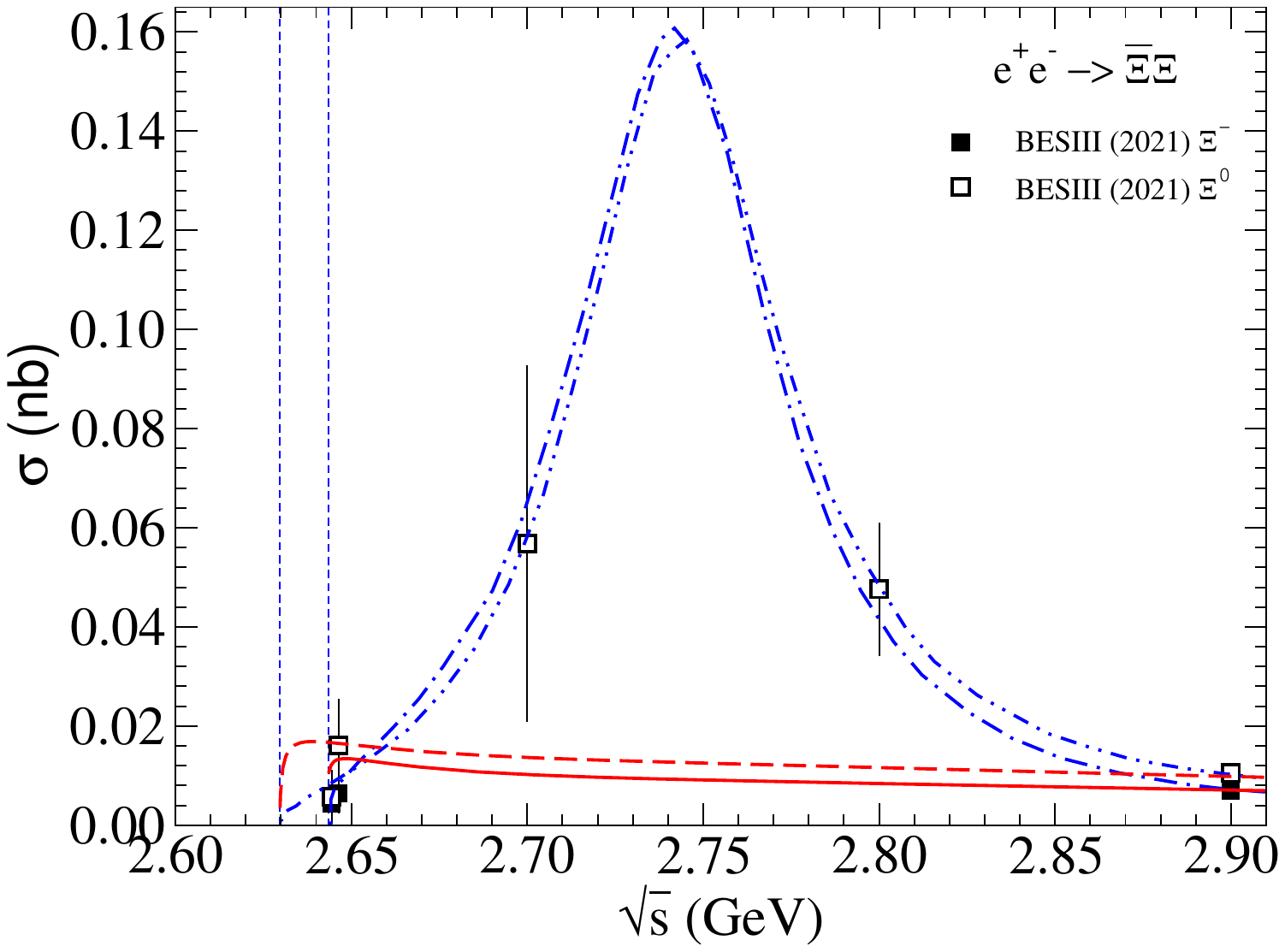}
\includegraphics[width=0.95\linewidth]{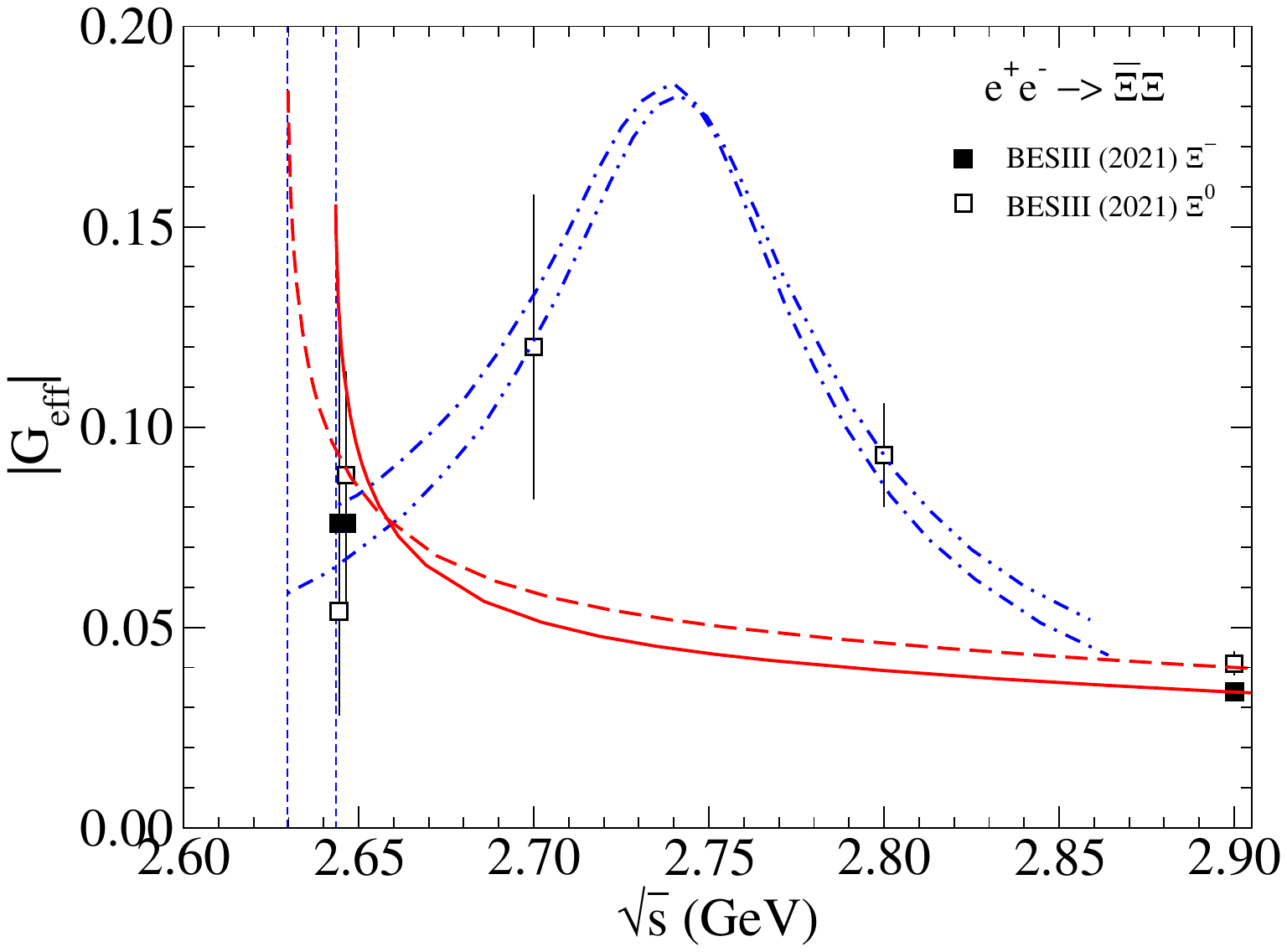}
\end{center}
\vskip -0.8cm 
\caption{Cross sections for $\ee\to\XXbz$ and $\ee\to\XXbm$. 
The data are from \cite{BESIII:2020ktn}
and \cite{BESIII:2021aer}.
Results based on $\XXb$ FSI effects~\cite{Haidenbauer:2020wyp} 
are shown by the solid and dashed lines. 
The dash-dotted (dash-double-dotted) line is the VMD result from
\cite{Yan:2023yff}. 
Vertical lines indicate the $\XXb$ thresholds. 
}
\label{fig:XX}
\end{figure} 

\sectitle{$\ee\to\XXb$ and the EMFFs of the $\Xi$}
Near-threshold measurements for $\ee \to \XXbm$ have been 
published by BESIII in 2020 \cite{BESIII:2020ktn} and soon 
afterwards also cross sections for $\XXbz$~\cite{BESIII:2021aer}. 
Ref.~\cite{Haidenbauer:2020wyp} included already some predictions 
for the reaction $\ee\to\XXb$, based on a $\XXb$ FSI 
\cite{Haidenbauer:1993wca}
afflicted by similar shortcomings as those for $\LSb$ and 
$\SSb$, see above. 
Since at the time when the calculation was performed
only experimental information on $\ee \to \XXbm$ 
was available~\cite{BESIII:2020ktn},  
the relative magnitude of the $\XXbm$ and 
$\XXbz$ channels could not be fixed, and only 
results in the isospin channels were shown. 
Now, results for both charge channels can be established, and those 
are presented in Fig.~\ref{fig:XX} together with the data  
for $\ee\to \XXbm, \XXbz$. Thereby we assumed
that the cross section for $\XXbz$ is slightly 
larger than the one for $\XXbm$, as indicated
by the data points at $2.9$~GeV. It should be said that the 
empirical cross section ratio is practically compatible 
with unity when considered over a larger energy range
\cite{BESIII:2021aer}.

An investigation of the reaction $\ee\to\XXb$ in the VMD approach 
can be found
in Ref.~\cite{Yan:2023yff}. Here in addition to $\rho$, $\omega$, 
and $\phi$ two broad poles at $2.742$ and $2.993$~MeV have been
added. Those results are reproduced in Fig.~\ref{fig:XX},
cf. the dash-dotted and dash-double-dotted lines. 
Obviously, with those vector mesons one can reproduce all data
points in the displayed energy region perfectly.
Nonetheless, it would be good to perform further experiments in 
order to reliably confirm or disprove a resonance-like behavior. 

\sectitle{$\ee\to\OOb$ and the EMFFs of the $\Omega$}
First measurements of the reaction $\ee\to\OOb$ had been 
reported already 10 years ago by CLEO \cite{Dobbs:2014ifa}, 
while more extended measurements have 
been performed only recently~\cite{BESIII:2022kzc}. 
However, the cross section turned out to be practically 
compatible with zero within the achieved accuracy. 
Moreover, there are no data close to the $\OOb$ threshold. 
A theoretical interpretation of the CLEO data has been attempted 
in \cite{Ramalho:2019koj,Ramalho:2020laj} within a pQCD
inspired model.

\begin{figure}[htbp] 
\begin{center}
\includegraphics[width=0.95\linewidth]{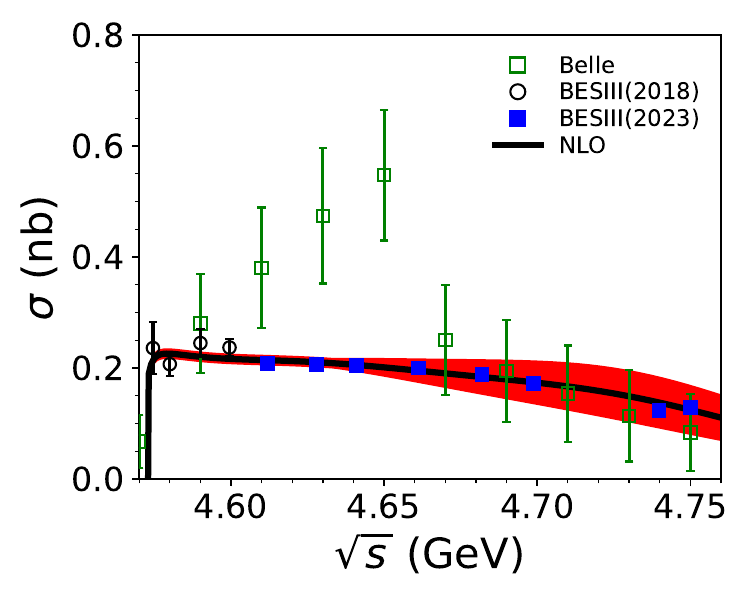}
\includegraphics[width=0.95\linewidth,height=0.6\linewidth]{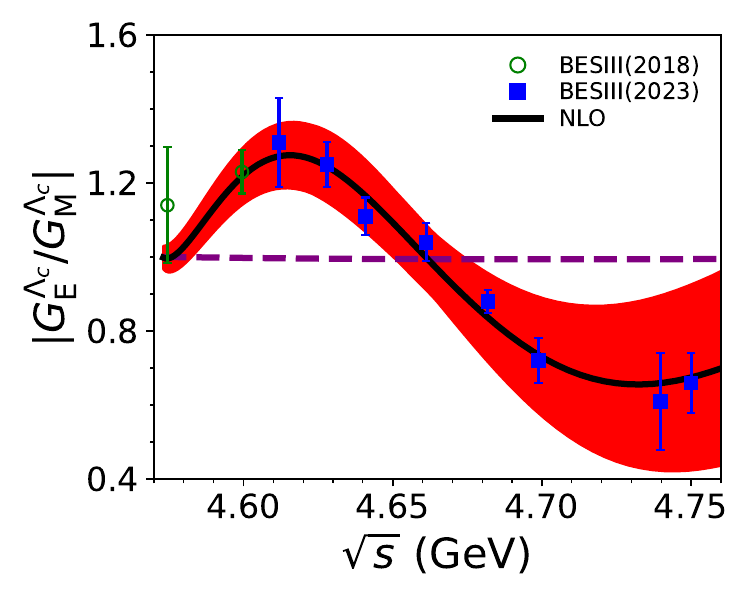}
\end{center}
\vskip -0.8cm
\caption{Cross section for $\ee\to\LcLcb$ and
ratio $|G_E/G_M|$.
Data are from Belle~\cite{Belle:2007umv}
and BESIII~\cite{BESIII:2017kqg,BESIII:2023rwv}. 
The theory results are by Guo et al.~\cite{Guo:2024pti}
based on an NLO $\LcLcb$ FSI.
}
\label{fig:Lc}
\end{figure}

\sectitle{$\ee\to\LcLcb$ and the EMFFs of the $\Lambda_c$}
Initial measurements of the $\ee\to\LcLcb$ cross section by the 
Belle Collaboration~\cite{Belle:2008xmh} indicated the presence
of a resonance not too far from the threshold, named $X(4630)$,
see Fig.~\ref{fig:Lc}.
Since a resonance with similar mass, the $Y(4660)$, 
had been established in other reactions, like
$\ee\to \pi^+\pi^-\psi(2S)$ \cite{Belle:2007umv}, 
there had been speculations from the very beginning, 
that one sees the same resonance in the two channels
\cite{Guo:2010tk,Dai:2017fwx}. 
However, the experiment by the BESIII Collaboration from 
2017~\cite{BESIII:2017kqg} signaled a possibly different behavior 
of the cross section, and the latest measurement 
by them \cite{BESIII:2023rwv} practically excludes the presence of 
a resonance. Indeed, now the cross section is more or less constant 
in the whole threshold region, a behavior quite 
similar to that observed for $\ee\to\LLb$ and $\ee\to\LSb$+c.c.. 
A theoretical interpretation of the new data has been accomplished 
in~\cite{Guo:2024pti}, based on a $\LcLcb$ FSI constructed
within chiral effective field theory up to next-to-leading order
(NLO). 
Their results for the cross section are included in Fig.~\ref{fig:Lc},
the ones for $G_{E}$ and $G_M$ can be found in~\cite{Guo:2024pti}.
Another interpretation of the data in terms of FSI effects can
be found in~\cite{Milstein:2022bfg,Salnikov:2023qnn}. 

Regarding analyses of $\ee\to\LcLcb$ within the VMD approach
we want to point to the works by Chen et al.~\cite{Chen:2023oqs,Chen:2024luh} where
 four charmonium-like states called 
 $\psi(4500)$, $\psi(4660)$, $\psi(4790)$, and $\psi(4900)$ have
 been included. 
 Earlier studies within the VMD model often focused on the now obsolete $X(4630)$. See for example~\cite{Wan:2021ncg}.

Interestingly, there is an oscillation in the ratio 
$|G_E/G_M|$ as measured by BESIII~\cite{BESIII:2023rwv}. 
This feature can be reproduced by
considering the $\LcLcb$ FSI, see Fig.~\ref{fig:Lc}, 
but also in the VMD approach \cite{Chen:2023oqs}. 

\sectitle{Summary}\label{Sec:IV}
In this Letter, we reviewed the present knowledge of the 
EMFFs of hyperons in the timelike region. As reported, 
regarding $\ee\to\YYb$ cross sections and/or effective
form factors, there is already a wealth of data on $\LLb$ and 
there has been an impressive increase in the database 
for other hyperon channels over the past few years. 
In general, theoretical studies can describe the energy dependence 
of the cross sections (or $G_{\rm eff}$) reasonably well, in the 
threshold region by taking into account the $\BBb$ FSI, or over a 
larger energy range within the conventional VMD model. 

On the other hand, 
for differential observables and/or the individual EMFFs 
$G_E$ and $G_M$ and their relative phase pertinent  
studies are still in their infancy. 
For the few channels/energies 
where data have been published, there is a discrepancy between
experiment and theoretical expectations. Moreover, with
just few data points available, the energy dependence of the 
quantities in question cannot be established reliably. 

Thus, 
it is perhaps fair to say that, at present the main unresolved 
problems and challenges are on the experimental side. 
For example, there are tensions between 
some experiments, notably in some $\SSb$ channels. 
Also, more cross-section data (and with better resolution) are required 
to firmly establish the energy dependence over the first $100$~MeV or so 
from the thresholds, where effects from the FSI are expected to play an
important role. This concerns in particular the $\LLb$ and $\LcLcb$
systems
with possibly a peculiar near-threshold behavior.  
And, of course, more differential cross sections and polarization data
are needed for a separation of $G_E$ and $G_M$ and for determining 
their relative phase. 

Finally, in recent times, there has been a rather strong interest in the 
apparent oscillations of the $G_{\rm eff}$'s of the proton and neutron 
in the timelike region \cite{Bianconi:2015owa}, 
see \cite{Yang:2022qoy} for an extensive list of references. 
Whether the EMFFs of hyperons exhibit such oscillations too has been addressed 
in Ref.~\cite{Dai:2021yqr} for the $\Lambda$, $\Sigma^0$, and $\Xi^0$.
It has been found that the present measurements are, in principle, 
qualitatively compatible with the presence of oscillations. 
However, due to the large experimental uncertainties in the 
pertinent data, no firm conclusion can be drawn at present.

\sectitle{Acknowledgements}
LYD acknowledges the support from the National Natural Science Foundation of China (NSFC) with Grants No.~12322502, 12335002, Joint Large Scale Scientific Facility Funds of the NSFC and Chinese Academy of Sciences (CAS) under Contract No.~U1932110, Hunan Provincial Natural Science Foundation with Grant No.~2024JJ3004, and Fundamental Research Funds for the central universities.
The work of UGM was supported by the CAS President's International
Fellowship Initiative (PIFI) (Grant No.~2025PD0022) and by the MKW NRW under the 
funding code No.~NW21-024-A and by ERC AdG EXOTIC
(grant No. 101018170).

\bibliographystyle{apsrev4-1}
\bibliography{ref}

\end{document}